\documentclass[10pt, conference]{IEEEtran}
\usepackage{graphicx}
\usepackage{times,amsmath,amsfonts}
\usepackage{amsmath,amssymb,amstext}
\usepackage{multirow}
\usepackage{hhline}
\usepackage{multirow}
\usepackage{cite}
\usepackage{epstopdf}
\usepackage{bm}
\usepackage{algorithm} 
\usepackage{algorithmicx}
\usepackage{algpseudocode}
\usepackage{color}
\usepackage[lofdepth,lotdepth]{subfig}
\usepackage{soul}
\usepackage{float}
\usepackage[letterpaper, left=0.7in, right=0.68in, bottom=1 in, top=0.75in]{geometry}
\usepackage{csquotes}
\usepackage{xspace}

\newcommand{\proposed}{\textsf{CAT3S}\xspace}

\newtheorem{proposition}{\textbf{Proposition}}

\begin{document}
%\title{\LARGE{Context-aware Spectrum Coexistence Design and Implementation in Satellite Bands - A Case Study of 12 GHz Band}}

\title{Context-Aware Spectrum Coexistence of Terrestrial Beyond 5G Networks in Satellite Bands}

%Context-Aware Terrestrial-Satellite Spectrum Sharing for Beyond 5G Wireless Networks}

%O-RAN Enabled Context-Aware Spectrum Coexistence of Terrestrial 5G in Satellite Bands

\markboth{IEEE International Symposium on Dynamic Spectrum Access Networks, 2024}%
{IEEE International Symposium on Dynamic Spectrum Access Networks, 2024}
%\author{Ta Seen Reaz Niloy, Rob Smith, Zoheb Hassan, Vikram R. Anapana, and Vijay K. Shah\\

\author{Ta Seen Reaz Niloy$^{\dagger}$, Zoheb Hasan$^{\dagger\dagger}$, Rob Smith$^{\dagger\dagger\dagger}$, Vikram R. Anapana$^{\dagger}$, and Vijay K. Shah$^{\dagger}$ \\
$^{\dagger}$\textit{NextG Lab@GMU}, George Mason University, Fairfax, VA, USA,\\ 
$^{\dagger\dagger}$ Universit\'e Laval, Qu\'ebec City, QC, Canada, and 
$^{\dagger\dagger\dagger}$The MITRE Corporation, USA \\
Emails:  \{tniloy, vanapana, vshah22\}@gmu.edu, rgsmith@mitre.org, md-zoheb.hassan@gel.ulaval.ca,}

\maketitle

\begin{abstract}
Spectrum sharing between terrestrial 5G and incumbent networks in the satellite bands presents a promising avenue to satisfy the ever-increasing bandwidth demand of the next-generation wireless networks. However, protecting incumbent operations from harmful interference poses a fundamental challenge in accommodating terrestrial broadband cellular networks in the satellite bands. State-of-the-art spectrum sharing policies usually consider several worst-case assumptions and ignore site-specific contextual factors in making spectrum sharing decisions, and thus, often results in under-utilization of the shared band for the secondary licensees. To address such limitations, this paper introduces \proposed ($\text{\underline{C}}$ontext-$\text{\underline{A}}$ware $\text{\underline{T}}$errestrial-$\text{\underline{S}}$atellite $\text{\underline{S}}$pectrum $\text{\underline{S}}$haring) framework that
%a dynamic spectrum sharing framework to facilitate coexistence between terrestrial 5G and incumbent networks over the satellite bands. \proposed 
empowers the coexisting terrestrial 5G network to maximize utilization of the shared satellite band without creating harmful interference to the incumbent links by exploiting the contextual factors. \proposed consists of the following two components: (i) \textit{context-acquisition unit} to collect and process essential contextual information for spectrum sharing and (ii) \textit{context-aware base station (BS) control unit} to optimize the set of operational BSs and their operation parameters (i.e., transmit power and active beams per sector). To evaluate the performance of the \proposed, a realistic spectrum coexistence case study over the 12 GHz band is considered. Experiment results demonstrate that the proposed \proposed achieves notably higher spectrum utilization than state-of-the-art spectrum sharing policies in different weather contexts.

\end{abstract}

% \begin{IEEEkeywords}
%    Context-awareness, dynamic spectrum sharing, radio resource optimization, interference management.
% \end{IEEEkeywords}

\vspace*{-0.1cm}
\section{Introduction}
The availability of sufficient spectrum is of paramount importance in enabling a diverse range of futuristic services, including connected cars, smart healthcare, remote surveillance, disaster management, environmental monitoring, and holographic communications, in the beyond 5G (B5G) cellular networks \cite{NGA_1}. However, the spectrum allocated to commercial 5G cellular networks is either overly congested, as in the FR1 band with sub-6 GHz frequency, or inherently limited in coverage, as in the FR2 band with 24 GHz or higher frequency \cite{mmWave_1}. In this context, sharing the bands primarily allocated to various non-terrestrial (i.e., satellite) communications presents a promising approach to alleviate the extensive spectrum demand of the B5G cellular networks. The wireless research community has recently shown interest in utilizing the mid-band spectrum for the B5G cellular networks, thanks to this band's ability to support long-range communication with high capacity \cite{Monisha}.  More specifically, the upper mid-band spectrum (7-24 GHz) is considered as the most vital band for deploying B5G networks by both regulators and standardization committee \cite{FCC_TAC, 3GPP}.  However, in various countries, including USA and Canada, such mid-band spectrum is primarily licensed to various non-mobile broadband communications (e.g., fixed satellite services (FSS)), government operations (radio-location), and scientific missions (e.g., weather satellite) \cite{Samsung}. These incumbent operations are critical and sensitive, and relocating these incumbents to alternative bands is challenging, expensive, and time-consuming.  Thus, \textit{a fundamental challenge of sharing (upper) mid-band spectrum with the B5G networks is to safeguard the incumbent operations from harmful interference while maximizing the utilization of shared spectrum for the coexisting cellular networks.}
\footnotetext{This paper has been accepted for publication in IEEE DySPAN 2024.}

% \setlength{\textfloatsep}{0pt}
% \begin{figure*}[t]
% \vspace{-0.3cm}
% \begin{center}		
%          \includegraphics[width=0.99\linewidth, draft=false]{direct_graphUmi_final.pdf}
% 		\caption{Boxplot of I/N ratio from individual MBS  for spectrum coexistence over the 12GHz band \cite[Fig. 3]{12GHz_WCL}.}
% 		\label{Fig_Intro_1}
% 		\vspace{-0.8cm}
% 	\end{center}
% \end{figure*}

% \setlength{\textfloatsep}{0pt}
% \begin{figure}[t]
% \begin{center}		
%          \includegraphics[width=1.00\linewidth, draft=false]{WCL_Fig_4_1.png}
% 		\caption{Active MBSs and aggregate I/N ratio for exclusion zone and deployment context-aware MBS activation policies for spectrum coexistence over the 12GHz band \cite{12GHz_WCL}.}
% 		\label{Fig_Intro_2}
%   \vspace{-0.2cm}
% 			\end{center}
% \end{figure}

State-of-the-art spectrum sharing policies typically use the worst-case assumption-based approaches to protect incumbents from harmful interference. For instance, the standard citizen broadband radio service (CBRS) in the USA over the 3.5 GHz band defines a predefined exclusion zone around the incumbents (i.e., navy radars) and deterministically turns off all the active radio links of secondary licenses within the predefined exclusion zone \cite{Reed}. Such a deterministic policy does not consider any contextual factors and thus, usually results in highly conservative spectrum utilization for the coexisting cellular networks. We emphasize that several contextual factors, such as environment (e.g., weather), network deployments (e.g., building and terrain profiles), base stations' (BSs') capabilities (e.g., beamforming and beam-nulling), satellite trajectory, incumbent receiver's location, and antenna pointing angle, strongly influence the resultant cellular-to-incumbent co-channel interference in spectrum sharing networks. By appropriately exploiting these contextual factors, utilization of shared spectrum in the coexisting cellular networks can be enhanced without creating harmful inference at the incumbent links \cite{12GHz_ComMag}. For instance, we have recently developed a realistic simulation-based spectrum coexistence study between the terrestrial 5G  and non-geostationary satellite orbit (NGSO) FSS links over the 12 GHz (12.2-12.7 GHz) band \cite{12GHz_WCL}. Such a study shows that by exploiting deployment contexts in activating cellular BSs, the utilization of the 12 GHz band for the coexisting terrestrial 5G networks is improved by 2.63 times. Accordingly, to facilitate the sharing of the satellite bands with the B5G cellular networks, \textit{context-awareness is of pivotal importance for the spectrum sharing policies.}

 For context-aware terrestrial-satellite spectrum sharing, a framework is required to (i) collect and process contextual information and (ii) optimize the available degrees of freedom according to these contexts. In various mid-band spectrums, the inherent directionality of terrestrial 5G BSs and incumbent receivers provides several degrees of freedom to optimize. For example, instead of completely turning off a BS, one can reduce its transmit power or deactivate a single sector. Similarly, different beams can be switched on/off at the BS(s) based on the incumbent antenna's pointing angle towards the satellite transmitter. In this work, we develop a spectrum-sharing framework to dynamically control the BSs' transmission parameters according to the networking contexts so that both co-channel interference is mitigated and spectrum utilization is improved. Note that our proposed framework is generic and, with no/little modifications, can apply to a wide range of satellite bands, including  $3.1-4.2$ GHz, $4.4-5$ GHz, $7.125-8.5$ GHz, and $12.7-13.25$ GHz band. 

\textbf{Contributions:} The specific contributions of this paper are summarized as follows. 

\smallskip \noindent $\bullet$  A novel context-aware spectrum sharing framework \proposed is developed for enabling spectrum coexistence of 5G broadband in satellite bands. \proposed comprises two key components --(i) a \textit{context-acquisition unit} to capture critical contexts for spectrum sharing, and (ii) \textit{context-aware BS control unit} that determines whether a particular BS is on/off, and if on, the optimal transmitter power and operational beam at each BS sector, such that the network throughput of the 5G network is maximized while maintaining the overall interference below an acceptable threshold at the incumbent receiver(s).

\smallskip \noindent $\bullet$ We model \textit{Context-aware BS control} as a resource optimization problem aimed at dynamically optimizing three key RF parameters of the coexisting 5G networks (i) set of operational 5G BSs in the shared satellite band, (ii) the active set of beams for each operation BS, and (iii) the operational transmit power of each operation BS. Such an optimization is provably NP-hard, and thus, a well-designed heuristic algorithm is proposed to obtain a sub-optimal yet computationally efficient solution. 
  
    %This algorithm is deployed as an xApp within the Near-RT RIC of the \proposed system.

\smallskip \noindent $\bullet$ For the case study, we consider spectrum sharing between downlinks of terrestrial 5G and NGSO FSS  over the $12$ GHz band in a semi-urban deployment scenario of Blacksburg, VA, and evaluate the \proposed's performance in different weather contexts. Notably, no real-world 5G deployments exist in the $12$ GHz band. Accordingly, we leverage and extend our recently developed 12 GHz interference analysis tool \cite{12GHz_WCL} to simulate the network environment and evaluate the  \proposed's performance. Experimental results show that \proposed simultaneously achieves a lower aggregate I/N ratio at the FSS receiver and higher downlink capacity and number of operational BSs compared to context-unaware spectrum sharing policies at all weather conditions.

\vspace{-0.2cm}
\section{Related Works}
Resource management is crucial in managing interference and improving spectrum utilization in the terrestrial-satellite spectrum-sharing networks \cite{Sat_Mobile_Sharing}. In \cite{Lit_Rev_1}, a detailed analysis of the interference resulting from the coexistence between 5G terrestrial and FSS networks over C-band (3.7-4.2 GHz) was presented, and interference mitigation strategies using switching on/off and backing-off transmit power of high-interference BS were proposed. Interference over the C-band can also be reduced by adjusting the exclusion zone's radius around the FSS receiver, transmission power, and beamforming pattern of the transmitting antenna \cite{Lit_Rev_2}. However, both \cite{Lit_Rev_1} and \cite{Lit_Rev_2} relied on heuristic mechanisms to control the networking parameters for interference mitigation. Concerning spectrum sharing over the S-band (3.55-3.65 GHz), a multi-tier dynamic protection zone was developed to protect primary incumbent users from interference and improve secondary users' performance in the CBRS system \cite{Lit_Rev_3}. Furthermore, a centralized channel allocation scheme was also developed to allocate channels among the paid and unlicensed users in the CBRS system while considering interference protection for incumbent and paid users \cite{Lit_Rev_4}. However, the studies mentioned above primarily considered static optimization and ignored dynamic variations of the channel and other parameters (e.g., user mobility) in the system. Dynamic resource optimization was investigated in the terrestrial-satellite integrated networks (TSINs), where both terrestrial and satellite users share the spectrum on a co-primary basis. In \cite{Lit_Rev_5}, spectrum sharing between uplink terrestrial and satellite users were studied, and a dynamic optimization for user-access point association, access link's bandwidth allocation, and users' transmit power allocation was developed. A delay quality-of-service aware transmit power allocation was proposed to maximize cognitive satellite network's downlink energy efficiency while keeping interference at the terrestrial networks smaller than a threshold \cite{Lit_Rev_6}. Considering the Ka-band sharing between satellite and terrestrial networks' downlink operations, hybrid beamforming, user scheduling, and power allocation were proposed to jointly maximize the sum throughput of the terrestrial and satellite networks \cite{Lit_Rev_7}. Recently, an open radio access network (O-RAN)-empowered framework was proposed to share spectrum between terrestrial cellular networks and government satellite systems, and a dynamic physical resource block (PRB) blanking scheme was developed to mitigate the interference from the terrestrial to satellite networks \cite{Lit_Rev_8}. However, context-aware resource optimizations are ignored in all these studies. Note that, unlike S and C bands, context-awareness is critically important for sharing upper mid-band spectrum between terrestrial cellular and incumbent networks\footnote{Please refer to \cite[Table 2]{12GHz_ComMag} for a detailed list of contexts that impact sharing performance of the upper mid-band spectrum.}. Essentially, the existing static and dynamic resource optimization schemes  \cite{Lit_Rev_1, Lit_Rev_2, Lit_Rev_3, Lit_Rev_4,Lit_Rev_5,Lit_Rev_6, Lit_Rev_7, Lit_Rev_8} are sub-optimal for upper mid-band spectrum sharing systems. 

In the context of USA, there is a significant interest from different stakeholders in opening the 12.2-12.7 GHz mid-band spectrum, commonly known as the 12 GHz band, for 5G terrestrial networks, thanks to the band's favorable propagation characteristics and contiguous 500 MHz bandwidth for both downlink and uplink communications \cite{Monisha_1}. In the USA, the 12 GHz band is primarily licensed to direct broadcast satellite (DBS) and NGSO FSS services, and because of their ubiquitous deployments, interference protection for such incumbent receivers is a key challenge in sharing the 12 GHz band. Some industry-specific studies investigated the interference between terrestrial 5G and incumbents in the 12 GHz band. For instance, \cite{DBS} studied interference between terrestrial 5G and DBS networks, while \cite{RKFreport,SpaceXanalysis} studied interference between terrestrial 5G and NGSO FSS networks. However, these studies adopted a probabilistic approach to interference evaluation and did not present any 5G network optimization schemes to mitigate interference. In \cite{12GHz_WCL}, we presented a realistic simulation-based interference analysis in the 12 GHz band and proposed heuristic schemes to select operational BSs to keep interference at the FSS receiver smaller than a threshold. However, such a study did not consider improving spectrum utilization for the coexisting 5G cellular networks.

\textbf{Uniqueness of the Current Work:} 5G multi-antenna BSs have the unique capability of creating beams in the desired directions and nulls elsewhere. Such a capability is instrumental in improving spectrum utilization of cellular networks and mitigating interference towards incumbent systems \cite{FCC_TAC}. Our current work develops a resource optimization framework to exploit 5G BSs' beamforming capability for activating a maximum number of BSs in a spectrum-sharing environment. More specifically, our proposed framework validates the suitability of the decision variables using a realistic interference evaluation framework, thereby taking different contextual factors in the BS's parameter optimization. Such context-awareness enables our proposed framework to activate more BSs and beams per BS's sectors compared to the existing approaches without harming the incumbent receivers. Consequently, our current work is unique compared to the state-of-the-art literature.

\vspace{-0.2cm}
\section{System Model}

\subsection{System Overview and Assumptions}
\begin{figure}
\vspace{-0.2cm}
 \begin{center}
		\includegraphics[width=0.95\linewidth, draft=false]{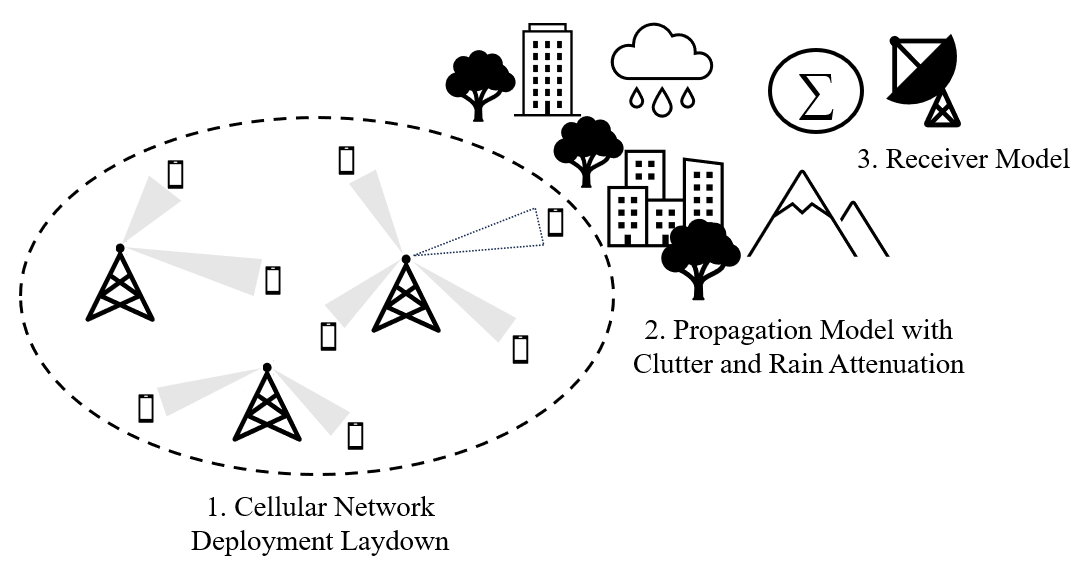}
		\caption{Overview of the proposed spectrum sharing scenario.}
		\label{fig_1}
	\end{center}
 \vspace{-0.2cm}
\end{figure}

Fig. \ref{fig_1} shows a terrestrial-satellite spectrum sharing scenario.  We assume that a total of $K$ 5G BSs are in the considered region, and $\mathcal{K}=\{1,2, \cdots, K\}$ is the set of all BSs. Each BS has three sectors, each covering a $120^\circ$ angular area. Each sector has a total of $M$ number of sub-arrays, each being fed with an RF chain, and there are a total of $N$ unique codebook vectors for each sub-array, each representing a unique 3D beam direction. The sets of sub-arrays and codebook vectors per sub-array are denoted by $\mathcal{M}$ and $\mathcal{N}$. Emissions from each sub-array propagate with path loss according to the deterministic path loss model, described in Section III.B.3. This model includes clutter and building information as well as rain attenuation. Interference from each active sub-array aggregates at the incumbent receiver, with the relative contribution from each sub-array affected by the antenna pattern and pointing angle of the sub-array, cellular BS's transmit power, path loss, and the antenna pattern and pointing angle of the FSS receiver. Detailed analysis of the interference generated from an active sub-array at the incumbent receiver is provided in the following sub-section. All the BSs are assumed to be connected with a centralized control center via a wired feedback line, and the centralized control center hosts a context-aware BS control unit of the \proposed framework, depicted in Fig. \ref{fig_2}, to dynamically adapt the set of active BSs and their RF parameters.

\setlength{\textfloatsep}{0pt}

For the tractability of the ensuing analysis, we make the following assumptions. \textbf{A1:} The incumbent receiver's location and antenna pointing angle information are available. \textbf{A2:} Channel hardening is achieved by the antenna arrays deployed at the cellular BSs. As such, the short-term channel fading is ignored, and only long-term channel impairments (i.e., path loss and shadow fading) are considered for both interference and access links. \textbf{A3:} Inter-cell interference in the access links are ignored since the cellular BSs typically optimize their electric down-tilt angles to minimize the inter-cell interference. At the same time, the intra-cell interference between two concurrently active beams at different sectors of a cellular BS is also considered negligible.  \textbf{A4:} The interference signal received from each sub-array sums constructively at the incumbent receiver. \textbf{A5:} In each cellular BS, only one beam per sector can be operational using the best possible transmit power and beam codebook, as obtained from the context-aware BS control unit of \proposed framework.
 
\vspace{-0.2cm}
\subsection{Analysis of Cellular-to-Incumbent Interference}
We develop an enhanced cellular-to-incumbent interference analysis tool by incorporating weather contexts and 5G-compliant codebook-based beamforming capability at the BSs into our previously developed interference analysis tool \cite{12GHz_WCL}. The components of the enhanced interference analysis tool are described in Section III.B.1-III.B.4, and the evaluation of interference from an active cellular BS to the incumbent receiver using this tool is provided in Section III.B.5.

\subsubsection{Transmitter Antenna Pattern of 5G BS}
5G BSs use a $M \times L$ Planar Array antenna for the codebook-based beamforming towards the UEs in each sector within its coverage area, where $M$ and $L$ are the numbers of rows and columns of the antenna element's array. The spacing between two consecutive antenna elements along the $x$ and $y$ axes is denoted as $d_x$ and $d_y$, respectively, and it is specified as $\lambda/2$ (where $\lambda$ is the wavelength). Progressive phase shifts are introduced as $\beta_x$ and $\beta_y$ the $x$ and $y$ axes, respectively. Each sector of a BS consists of $4$ antenna sub-arrays; thus, each BS contains a total of $12$ sub-arrays for all the $3$ sectors. Each sub-array is equipped with a total of $64$ predefined beam codebooks. For shaping the beams by taking into account the antenna array configuration, antenna elements spacing, and antenna array size, we leverage the planner antenna array factor \cite{Balanis}. For the given azimuth and elevation angles $\theta \in [0, \pi]$ and $\phi \in [-\pi, \pi]$, such an array factor is defined as
\begin{equation} \label{eq:euation_AF}
AF(\theta, \phi) = 
\left( \frac{1}{M}\frac{\sin^2(\frac{M}{2}\psi_{x})}{\sin^2(\frac{\psi_{x}}{2})}\right)
\left( \frac{1}{L}\frac{\sin^2(\frac{L}{2}\psi_{y})}{\sin^2(\frac{\psi_{y}}{2})}\right)
\end{equation}
Here, the phase differences  $\psi_{x}$ and $\psi_{y}$ for the $k$-th wave number are expressed as $\psi_{x} = kd_{x}\sin\theta \cos\phi + \beta_{x}$ and $\psi_{y} = kd_{y} \sin\theta \cos\phi+ \beta_{y}$, respectively. The overall antenna gain of $i$-th beam (in dBi) transmitted from the the $m$-th sub-array of the $k$-th BS at ($\theta, \phi$) angle is obtained as
\begin{equation} 
\label{eq:euation_gain}
G_{5G}^{(i)}(\theta, \phi) = 10\log_{10}[AF(\theta, \phi)].
\end{equation}

\subsubsection{Receiver Antenna Pattern of Incumbent}
The Class B wide band earth stations (WBES) antenna pattern, standardized by ETSI, is considered for the incumbent receiver. In particular, for the 12 GHz case study in Section VI,  antenna gain for the incumbent (FSS) receiver at $\phi$ angle relative to the boresight direction is computed using \cite[eq. (8)]{12GHz_WCL}.   
% \begin{equation} 
% \label{eq:IN_equation_fss}
%    G_{SAT}(\phi) =
%    \begin{cases}
%    40- 25 \log \phi ~ \text{dBi} & 6^{\circ} \leq \phi < 48^{\circ}\\
%    -2 ~  \text{dBi} & 48^{\circ} \leq  \phi \leq 180 ^{\circ}.
%    \end{cases}
%    \end{equation}

\subsubsection{Path Loss Model}
Unlike the probabilistic method of computing path loss \cite[Table 7.4.2]{pathloss}, we develop a deterministic path loss model by accurately determining line-of-sight (LOS) and non-LOS (NLOS) propagation paths between the cellular BSs and incumbent receivers. This model considers various site-specific contextual factors, including geographic locations of the BSs and incumbent receivers, signal obstruction caused by buildings, and weather-induced signal attenuation. We first incorporate the required building information, including buildings' heights, geographical coordinates, and shapes, obtained from the OpenStreetMap database \cite{OpenStreetMap} within the considered region. Subsequently, we determine whether any building polygons intersect the interference axis between a cellular BS and an incumbent receiver. The link between a cellular BS and incumbent receiver is classified as NLOS and LOS, respectively, if there are one/more and no intersections. Considering the distance between the cellular BS and FSS receiver is $d_k$ (in km), the path loss is computed by
\cite{12GHz_WCL}.
 \begin{equation}
     \label{PL}
     \begin{split}
    &     \text{PL}(d_k)= \mathbf{1}_{(\beta=0)}\left(\text{PL}_{\mbox{\tiny{NLOS}}}(d_k)+X(\sigma_{\mbox{\tiny{NLOS}}})\right)\\
  &  + \mathbf{1}_{(\beta=1)}\left(\text{PL}_{\mbox{\tiny{LOS}}}(d_k)+X(\sigma_{\mbox{\tiny{LOS}}})\right)\
     \end{split}
 \end{equation}
where  $\mathbf{1}_{(\cdot)}$ is an indicator function; $\beta=0$ and $\beta=1$ indicate the existence of NLOS and LOS signal propagation paths, respectively;  $X(\sigma_{\mbox{\tiny{b}}})$ indicates shadow fading loos (in dB) with $\sigma_b$ as the standard deviation for $b \in \{\mbox{LOS, NLOS}\}$; $\text{PL}_{\mbox{\tiny{LOS}}}(\cdot)$ and $\text{PL}_{\mbox{\tiny{NLOS}}}(\cdot)$ indicate path loss (in dB) over LOS and NLOS propagation paths, respectively. Without loss of generality, we leverage the 3GPP path loss models, \cite[Table 7.4.1]{pathloss} for path losses over LOS and NLOS links.

Eq. \eqref{PL} provides exact path loss in the clear (i.e., sunny)  weather conditions. Note that the satellite signals of the upper mid-band spectrum exhibit several attenuation in rainy weather. To account for that, inspired from \cite{PL_Rain}, we incorporate an additional attenuation factor in the \eqref{PL}. Accordingly, the path loss in the rainy weather condition is obtained as 
 \begin{equation}
    \label{PLW}
    \text{PL}(d_k)_{(Rainy)} =\text{PL}(d_k)+ \text{A}_{Rain} \times d_k
    \end{equation}
where $\text{A}_{Rain}$ is the rain-induced attenuation factor in dB/km unit. Using a mathematical model  on Rain Wave Attenuation due to Rain (RWAR) \cite{PL_Rain}, which is valid for both vertical and horizontal polarization for 10-100 GHz of the frequency range, the rain-induced attenuation factor is expressed as
\begin{equation}
    \label{AF}
    \begin{split}
    \text{A}_{Rain}= af^3+ bf^2+ cf+d
    \end{split}
    \end{equation}
where $a$, $b$, $c$, and $d$ are rain rate-dependent constants. Leveraging a curve fitting algorithm for vertical polarization \cite{PL_Rain} and extracting rain rate in mm/h unit (denoted by $x$) from the OpenWeatherMap API\cite{OpenWeathertMap}\footnote{OpenWeather API provides weather information of a specific day for a certain geolocation or city}, these constants are expressed as
\begin{equation}
     \label{a}
     \begin{split}
     a = -5.520\times10^{-12} x^3 + 3.26\times10^{-9} x^2\\
     - 1.21x\times 10^{-7} - 6\times 10^{-6}, 
     \end{split}
     \end{equation}
 \begin{equation}
     \label{b}
     \begin{split}
     b = 8\times10^{-10} x^3 - 4.522\times10^{-7} x^2\\ 
     - 3.03x\times 10^{-5}+ 0.001,
     \end{split}
     \end{equation}
     \begin{equation}
     \label{c}
     \begin{split}
     c = -5.71\times10^{-9} x^3 + 6\times10^{-7} x^2\\
     +8.707x\times 10^{-3}- 0.018,
     \end{split}
     \end{equation}
     and
 \begin{equation}
     \label{d}
     \begin{split}
     d = -1.073\times10^{-7} x^3 + 1.068\times10^{-4} x^2\\
     -0.0598x\times 10^{-3} +0.0442.
     \end{split}
     \end{equation}

\subsubsection{Impact of the Incumbent's and Secondary Link's Channel Usage} 
The shared band usually has multiple  channels. 
%Hence, the incumbent's and secondary link's activity over these channels impact the co-channel interference. Notably, the incumbent and secondary links of a given geographic region may not necessary always use the same channels. 
As a result, a beam transmitted from a cellular BS interferes with the incumbent receiver if and only if such a beam is operational in a channel that is under use at the incumbent. 
%Meanwhile, in a cellular BS, a beam cannot be operational in a channel if  no user equipment (UE) is scheduled  to this channel. These factors need to carefully considered in the interference analysis. 
To determine whether a beam will interfere with the incumbent link or not,  we introduce an binary indicator $\sigma_{k,j} \in (0,1)$ for the $j$-th sector ($j \in  \{1,2,3\}$) of the $k$-the BS, where $\sigma_{k,j} =1$ implies that the beam transmitted in the $j$-th sector of the $k$-th cellular BS interferes with the incumbent and $\sigma_{k,j}=0$ otherwise. This variable can be directly determined from the real-time channel-usage  and user scheduling information at the incumbent receiver and cellular BSs, respectively. More precisely, $\sigma_{k,j} =1$ if some users in the $j$-th sector of the $k$-th cellular BS are scheduled to the channel used by incumbent and $\sigma_{k,j} =0$ otherwise.

\subsubsection{Evaluation of Cellular-to-Incumbent Interference} 
Here, we compute the interference received from the $k$-th cellular BS to the incumbent receiver. Without loss of generality, we consider that at the $j$-th sector of the $k$-th BS, where $j \in \{1,2,3\}$, the $n_j$-th directional beam is transmitted from the $m_j$-th sub-array,
%\footnote{The active BSs, their transmit power, and the active beams in each sector of an active BS are determined using the proposed Algorithm 1.}, 
and the total transmit power of the $k$-th BS is equally divided among the sectors. 
Accordingly, the co-channel interference (in dB) introduced at the incumbent receiver by the $j$-th sector of the $k$-th BS is expressed as
\begin{equation}
    \label{interference_sector}
    \begin{split}
    &I_{k,m_j,n_j}^{(j)}= 10\log_{10}(P_k)+ G_{5G}^{(k)}\left(\theta_{n_j,int}, \phi_{n_j,int}\right)\\
    &+G_{SAT}(\phi_{k, SAT})-\text{PL}(d_{k,SAT})-10\log_{10}(3)\\
\end{split}
\end{equation}
where $P_k$ is the transmit power of the $k$-th cellular BS; $\theta_{n_j,int}$ and $\phi_{n_j,int}$ are the azimuth and elevation angles between the interference axis (from the $k$-th BS to the incumbent receiver) and the $n_j$-th beam, respectively;  $\phi_{k, SAT}$ is the angle between incumbent receiver's boresight direction and interference axis from the $k$-th BS;  $d_{k, SAT}$ is the distance between the $k$-th BS and incumbent receiver.  Thus, the total interference received from  the $k$-th BS to the incumbent receiver is expressed as
\begin{equation}
    \label{interfernece}
    I_k= \sum_{j=1}^3 \sigma_{k,j} 10^{\frac{I_{k,m_j,n_j}^{(j)}}{10}}.
\end{equation}

\section{\proposed Framework}
\begin{figure}
\vspace{-0.2cm}
	\begin{center}
		\includegraphics[width=1\linewidth, draft=false]{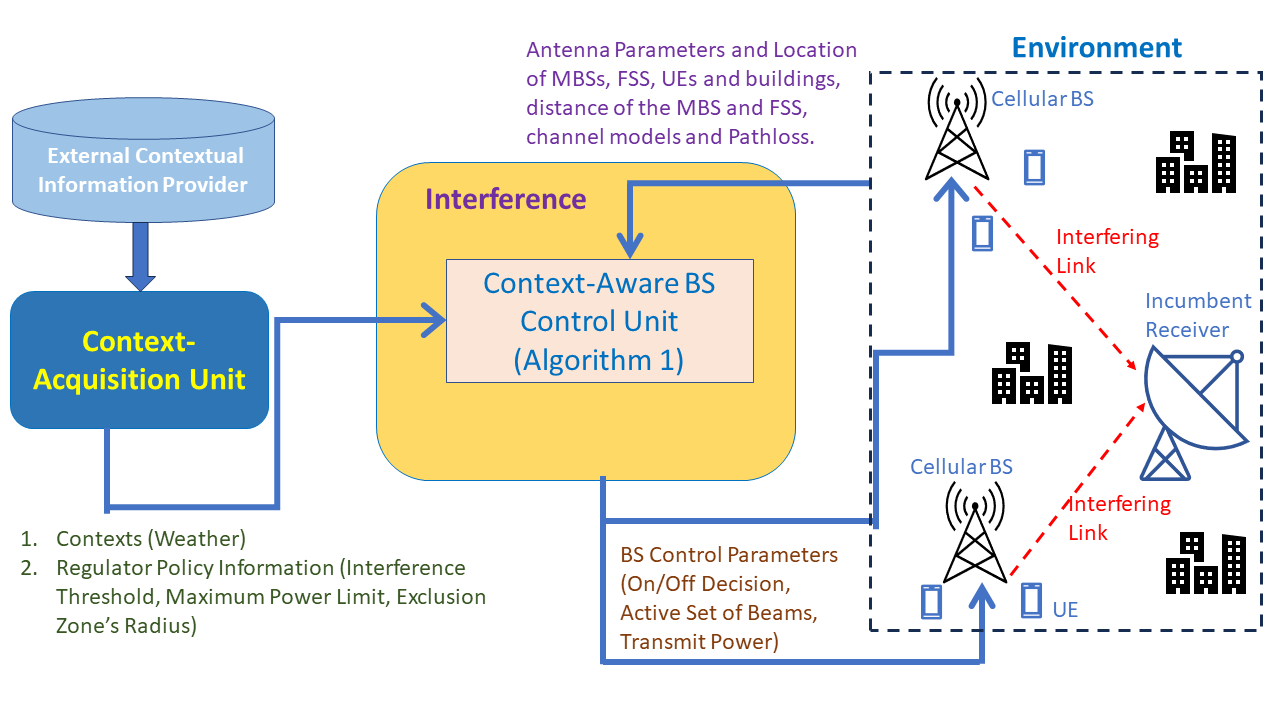}
		\caption{Overview of \proposed framework}
		\label{fig_2}
	\end{center}
 \vspace{-0.2cm}
\end{figure}

Figure \ref{fig_2} comprehensively depicts the \proposed framework with all the essential units.  \proposed comprises of the following three key components.

%\begin{itemize}
\smallskip \noindent $\bullet$ \textbf{Context-acquisition unit:} This unit gathers critical context information for spectrum sharing, such as weather data (e.g., rain data), interference thresholds for different weather conditions, exclusion zone radii, and different transmitting power levels of BSs, from the external/internal context information providers.

\smallskip \noindent $\bullet$ \textbf{Context-aware BS Control Unit:} This unit implements Algorithm \ref{Algorithm1}, detailed in the Section IV. B, to dynamically select the following parameters of the coexisting cellular network, namely, (i) the set of operational 5G BSs in the shared satellite band, (ii) the active set of beams for each operational BS, and (iii) the operating transmit power of each operational BS.

\smallskip \noindent $\bullet$ \textbf{Environment:} It refers to the terrestrial 5G-satellite spectrum sharing network. Due to the lack of real-world 5G deployments in the satellite band, we exploit our developed interference analysis tool to mimic the spectrum sharing environment. Note that this tool integrates information on the realistic 5G BSs deployments, satellite receiver's positions and antenna pointing angles, detailed 3D building information (geo-locations, heights, footprints, and polygon shapes), and user locations. As a result, this tool helps to infer site-specific path losses and interference caused by a set of BS parameters.
%\end{itemize}
%The proposed framework comprises three key components,  the context-acquisition unit, the realistic interference analysis environment, and the context-aware BS control unit.  . 
%The context-acquisition unit gathers critical context information, including weather data (rain rate or clear weather), interference thresholds for different weather conditions, exclusion zone radii, and different transmitting power levels.
%Subsequently, the interference analysis environment integrates information on the 5G MBSs deployment, FSS receiver's positions, building geolocations, and user locations. For detailed and ease of analysis, we employ an interference analysis tool for the evaluation of interference to noise ratio considering different weather conditions, FSS pointing angles, and various transmitting antenna patterns. This tool helps to calculate site-specific path loss considering distances between FSS, MBSs, and UEs, alongside parameters like 5G MBS transmitter gain, FSS receiver gain, and channel models.

 \proposed implements the following three steps. In Step I, the context-acquisition unit and environment provide feedback on the acquired contexts and received interference from the coexisting cellular network, respectively. In Step II, the context-aware BS control unit executes Algorithm \ref{Algorithm1} while considering such received information. In Step III, the final output parameters obtained from the context-aware BS control unit are provided to the BSs over reliable feedback channels, and BSs perform downlink data transmission using these parameters such that the total downlink capacity is increased without creating harmful interference to the incumbent link.

%Leveraging the data gathered from both the Interference Analysis Environment and the Context Acquisition unit, the Context-Aware BS Control Unit initiates its operations to determine the status of MBSs. It determines the active set of MBSs, their corresponding beams, and their associated transmitting power levels. In the following sections, a detailed description of each component is provided.

\setlength{\textfloatsep}{0pt}

%%%%context-acquisition and context-aware BS control. A nice detailed figure as well as walk through of the entire framework is required.

%%%%% We need the interference-evaluation tool as well to estimate the aggregate interference in a trustworthy manner as such this interference can be used in the control algorithm. The key motivation behind this that without such a framework, we need to have a reliable feedback channel from the incumbent to controller to feedback the real-time interference measurent report, every time the control decision is updated, which increases overhead/latency. In this work, we consider that this interference evaluation tool is exactly same to the tool provided in Section III. B. One may argue that the we may not have the perfect information of network or dynamic context, and hence, the estimated interference can be different than the exact one. To mitigate that, we may design DL-enabled interference eavluation tool  to predict the exact interference from the estimated interference. However, this is left for the future extension.

%\vspace{-0.2cm}
\section{Context-Aware BS Control Problem Formulation and Algorithm Development}
 
 In this section, we develop an optimization problem to dynamically control BS's parameters (Section V. A) and an algorithm to solve such a problem (Section V. B).
 %algorithm to dynamically control BS's parameters to enhance the shared band utilization at the coexisting networks without harming incumbent operations. Sections V. A and V. B  formulates an RF parameter optimization problem and develops an efficient control algorithm, respectively
 % three key RF parameters of the coexisting cellular networks, namely (i) the set of operational 5G BSs in the shared satellite band, (ii) the active set of beams for each operational BS, and (iii) the operating transmit power of each operational BS, .  %Note that this work does not consider availability of the real-time channel usage information of the incumbent. Essentially, we consider the \textit{worst-case scenario} by assuming that the incumbent  \textit{uses the entire available bandwidth}, and thereby, regardless of scheduling of UEs to the BS's channels, the transmitted beam(s) always interfere with the incumbents. The  current BS control framework's capability can be enhanced by sensing the incumbent's real-time channel usage (possibly using a sensing mechanism \cite{Sense_ORAN}) and optimizing UE scheduling at the cellular BSs as such the transmitted beams only use the channels not occupied by the incumbent.  Such an extension will be considered in our future works. 

\vspace{-0.3cm}
\subsection{Optimization Problem Formulation}
We first introduce the optimization variables.  Let $\nu_k \in \{0,1\}$ be the BS activation indicator where $\nu_k=1$ implies that the $k$-th BS is operational in the shared satellite band, and otherwise, $\nu_k=0$. We also introduce another binary variable $z_{k,m,n}^{(j)} \in \{0,1\}$ where $z_{k,m,n}^{(j)}=1$ implies that the $n$-th codebook vector is loaded to the $m$-th sub-array at the  $j$-th sector of the $k$-th BS and otherwise, $z_{k,m,n}^{(j)}=0$. Let $F_{k,m,n}^{(j)}$ be the total number of UEs supported by the beam created by loading the $n$-th codebook vector to the $m$-th sub-array at the  $j$-th sector of the $k$-th BS. The received signal-to-noise-ratio (SNR) at the $u$-th UE, $\forall u \in \{1,2, \cdots, F_{k,m,n}^{(j)}\}$, is expressed as
\begin{equation}
    \label{SINR_u_k}
    \begin{split}
    \text{SNR}_{k,m,n}^{(u,j)}= & 10\log_{10}(P_k)+ H^{(m,n)}\left(\Theta_{n}, \Theta_u\right)-\text{PL}(d_{k,u})\\
    &-10\log_{10}(NP)-10\log_{10}(3)
\end{split}
\end{equation}
where  $H^{(m,n)} (\Theta_{n}, \Theta_u)$ is the overall antenna gain (in dBi) from the $n$-th beam towards the $u$-th UE with $\Theta_n$ and  $\Theta_u$ as the angle-of-departure and angle-of-arrival, respectively; $\text{PL}(d_{k,u})$ is the path loss (in dB) with $d_{k,u}$ as the distance between the $k$-th BS and $u$-th UE; and $\text{NP}$ is the noise power in Watt. The achievable downlink capacity (in bps/Hz unit) at the $u$-th UE  is obtained as $C_{k,m,n}^{(u,j)}=\log_2\left(1+10^{SNR_{k,m,n}^{(u,j)}/10}\right)$. Therefore, the total achievable downlink capacity of the $k$-th BS is obtained as
\begin{equation}
    \label{capacity}
    C_k=\sum_{j=1}^3\sum_{m=1}^M\sum_{n=1}^{N} \sum_{u=1}^{F_{k,m,n}^{(j)}} z_{k,m,n}^{(j)} C_{k,m,n}^{(u,j)}.
\end{equation}

A resource optimization problem to maximize the utilization of the satellite band for the coexisting 5G cellular network while ensuring interference protection for incumbent satellite receivers is formulated as $\text{P0}$ at the top of the next page.
\begin{table*}
    \vspace{-0.5cm}
        \begin{normalsize}
\begin{equation}
	\label{optimization}
	\begin{split}
		\text{P0:} & \max_{\substack{ \bm{\nu} \in \{0,1\}, \\ \bm{z} \in \{0,1\}, \mathbf{P}}}  (1-w)\sum_{k=1}^K \nu_k C_k+ w\sum_{k=1}^K \sum_{j=1}^3 \sum_{m=1}^M\sum_{n=1}^{N} \nu_k z_{k,m,n}^{(j)} F_{k,m,n}^{(j)}\\
			& \text{s.t.} \begin{cases}
					&\hspace*{-0.3cm}  \text{C1:}~ \hspace{0.2cm} \sum_{k=1}^K \sum_{j=1}^3 \sum_{m=1}^M\sum_{n=1}^{N} \nu_k z_{k,m,n}^{(j)}  10^{\frac{I_{k,m,n}^{(j)}}{10}} \leq I_{th}\\
                        &\hspace*{-0.3cm}  \text{C2:}~ \hspace{0.2cm} \min_{u \in \mathcal{U}_{k,j}} \sum_{m=1}^M \sum_{n=1}^N z_{k,m,n}^{(j)} C_{k,m,n}^{(u,j)} \geq R_{th}, \forall j \in \{1,2,3\}, k  \in \mathcal{K}\\
                        &\hspace*{-0.3cm} \text{C3:}~ P_k \leq P_{max}, \forall k \in \mathcal{K}\\
                        %&\hspace*{-0.3cm}  \text{C4:}~ r_e \geq r_{min}\\
                        &\hspace*{-0.3cm} \text{C4:} \sum_{m=1}^M\sum_{n=1}^N z_{k,m,n}^{(j)}=1, \forall j,k.\\    \end{cases}
	\end{split}
\end{equation}
\end{normalsize}
\hrulefill
\end{table*}
In $\text{P0}$, the objective function maximizes a weighted sum of the achievable downlink capacity of the 5G cellular network and the number of UEs it supports. Here, $w \in (0,1)$ is a predefined weight factor. The considered objective function maximizes the utilization of the shared satellite band for the coexisting 5G cellular network while ensuring fairness.  In $\text{P0}$, constraint $\text{C1}$ implies that the total interference introduced by the coexisting  5G cellular network to the incumbent receiver must be smaller than a predefined threshold $I_{th}$; constraint $\text{C2}$ implies a rate quality-of-service (QoS) constraint for the coexisting cellular networks'  UEs with $C_{th}$ as the minimum required downlink data rate and $\mathcal{U}_{k,j}$ as the set of all UEs in the $j$-th sector of the $k$-th BS; constraint $\text{C3}$ implies that the transmitted power from each BS must be smaller than a regulator defined maximum transmit power limit, $P_{max}$, over the shared satellite band; constraint $\text{C4}$ implies that a pair of one codebook vector and one sub-array are selected at each sector of a given BS, i.e., at most one beam can be operational in each sector of a given BS. 

\begin{proposition}
    $\text{P0}$ is an NP-hard optimization problem.
\end{proposition}

\proof It can be shown by reducing PO to a 0-1 Knapsack problem, a classic NP-hard problem. The proof is omitted due to space limitations.

Since $\text{P0}$ is an NP-hard optimization problem, it cannot be solved in polynomial time. In the next sub-section, we propose a sub-optimal yet computationally efficient algorithm to solve $\text{P0}$ in polynomial time.
\setlength{\textfloatsep}{0pt}
\begin{algorithm}
	\caption{\textbf{Context-Aware BS Control Algorithm}}
	\label{Algorithm1}
	\begin{algorithmic}[1]
 \scriptsize
     \State \textbf{Input:} Contextual factors, including weather, RF propagation model, 5G deployment information (location and heights of BSs and buildings in the environment), incumbent receiver's location and pointing angle; interference threshold $I_{th}$.

\State Determine the minimum transmit power $P_{min}^{(k)}$ for the $k$-th BS to satisfy the constraint $\text{C2}$, $\forall k \in \mathcal{K}$, and set $P_{min}=\max_{k \in \mathcal{K}} P_{min}^{(k)}$.

\State Determine the set of non-operational BSs  $\mathcal{K}_{off}=\{k \in \mathcal{K}| P_{min}^{(k)} >P_{max} \}$. Assign $\nu_k^*=0$, $z_{k,m,n}^{(j)^*}=0$, and $P_k^*=0$, $\forall k \in \mathcal{K}_{off}, j \in \{1,2,3\}, m \in \mathcal{M}, n \in \mathcal{N}$.

\State \textbf{Initialize:} The set of operational BSs $\mathcal{K}_o= \mathcal{K} \setminus \mathcal{K}_{off}$;  $\nu_k^*=0$, $z_{k,m,n}^{(j)^*}=0$, and $P_k^*=0$, $\forall k \in \mathcal{K}_o, j \in \{1,2,3\}, m \in \mathcal{M}, n \in \mathcal{N}$; $\mathbb{B}_{op}=\varnothing$; $I_{agg}=0$; Outer\_Iteration\_Index=1; Number\_of\_BS\_Power\_Change~=~0.

\For{ $P \in [P_{min}, P_{max}]$} ~~(\textbf{outer loop})

\State \textbf{Initialize:} $\mathbf{U}_{BS}= \mathcal{K}_o$, $\mathbf{U}_{sector}^{(k)}=\{1,2,3\}$, $P_k=P$, $\forall k \in \mathbf{U}_{BS}$;  

\State  Calculate the following score: 
\[\rho_{k,m,n}^{(j)}=\frac{ (1-w)\sum_{u=1}^{F_{k,m,n}^{(j)}} C_{k,m,n}^{(u,j)} +w F_{k,m,n}^{(j)}}{10^{\frac{I_{k,m,n}^{(j)}}{10}}}\], $\forall k \in \mathbf{U}_{BS}, j \in \mathbf{U}_{sector}^{(k)},   m \in \mathcal{M}, n \in \mathcal{N}$.

\While {$\mathbf{U}_{BS} \neq \varnothing $} ~~(\textbf{inner loop})
    \State Find $(k^*, j^*, m^*, n^*)=\arg \max_{\substack{k \in \mathbf{U}_{BS}, j \in \mathbf{U}_{sector}^{(k)} \\ m \in \mathcal{M},\\ n \in \mathcal{N}}} \rho_{k,m,n}^{(j)}$.
    
    \State Calculate $I_{agg}^{Temp}= I_{agg}+10^{\frac{I_{k,m,n}^{(j)}}{10}}$.
    
   \If {$I_{agg}^{Temp} > I_{th}$}

    \State \textbf{Break};

    \Else
    
        \If{Outer\_Iteration\_Index ~ == ~1}

           \State Assign $\mathbb{B}_{op}\leftarrow \mathbb{B}_{op} \cup (k^*, j^* m^*, n^*)$;  Update $\mathbf{U}_{sector}^{(k^*)} \leftarrow \mathbf{U}_{sector}^{(k^*)} \setminus \{j^*\}$ and $\mathbf{U}_{BS} \leftarrow\{ k \in \mathbf{U}_{BS}\bigg| |\mathbf{U}_{sector}^{(k)}|  \geq 1\}$.

            \State Assign $z_{k^*,m^*, n^*}^{(j^*)^*}=1$,  $\nu_{k^*}^*=1$, $P_{k^*}^*=P$, and $I_{agg} \leftarrow I_{agg}^{Temp}$.  

          \State Update  Number\_of\_BS\_Power\_Change++

        \Else
               \If {$\mathbb{B}_{op}$ is empty}

                \State Update $\mathbb{B}_{op}\leftarrow \mathbb{B}_{op} \cup (k^*, j^* m^*, n^*)$.

                \Else 

                \State Replace current indices of the sector, sub-array, and codebook of the $k^*$-th BS with the indices computed in the Step 9.

                \EndIf
                
        \State Update $\mathbf{U}_{sector}^{(k^*)} \leftarrow \mathbf{U}_{sector}^{(k^*)} \setminus \{j^*\}$ and $\mathbf{U}_{BS} \leftarrow\{ k \in \mathbf{U}_{BS}\bigg| |\mathbf{U}_{sector}^{(k)}|  \geq 1\}$.

        \State Assign $z_{k^*,m^*, n^*}^{(j^*)^*}=1$,  $P_{k^*}^*=P$, and $I_{agg} \leftarrow I_{agg}^{Temp}$.  

          \State Update  Number\_of\_BS\_Power\_Change++
        \EndIf
    
    \EndIf

 \EndWhile 

\State Update $\mathcal{K}_o=\left\{k \in \mathcal{K}_o \bigg|\nu_k^*=1\right\}$.

\If {Number\_of\_BS\_Power\_Change is not changed}

\State \textbf{Break};

\EndIf

\EndFor

\State \textbf{Output:} $\{\nu_k^*\}$, $\{z_{k,m,n}^{(j)^*}\}$, $\{P_k^*\}$.
\end{algorithmic}
\end{algorithm}
\vspace{-0.15cm}
\subsection{Proposed BS Control Algorithm}
We propose a context-aware  BS control algorithm, summarized as Algorithm  \ref{Algorithm1}, to
efficiently solve $\text{P0}$. Algorithm \ref{Algorithm1} iteratively optimizes the beam-domain and power-domain control tasks in the inner and outer loops, respectively. More specifically, the beam-domain control selects the active BSs and suitable beams in the active sectors for the given transmit power. Meanwhile, the power-domain control searches for suitable transmit power of the BSs within a feasible range. In the first iteration, the inner loop activates a set of BSs and selects beams in their sectors assuming that all BSs operate at the minimum possible transmit power. The primary objective here is to activate the maximum number of BSs within the shared band without violating interference constraint $\text{C1}$. However, this approach still results in a conservative spectrum sharing scenario because each active BS is allowed to use only the minimum power level. In practice, depending on contextual factors such as weather and deployment scenarios, some BSs can transmit at high power levels without causing harmful interference to incumbent satellite receivers while selecting suitable beam directions. Thus, in the subsequent iterations, the outer loop gradually increases transmit power, while the inner loop further optimizes beam steering for the active BSs for the selected power. This allows some BSs to operate using high transmit powers in the shared band while keeping the aggregated interference at the incumbent below a tolerable threshold. It's worth noting that the set of active BSs remains unchanged after the first iteration; the only variables that change are the active BSs' transmit powers and the beams within their sectors.

The key steps of Algorithm \ref{Algorithm1} are explained as follows. Step 2 determines the minimum value of the transmit power for all the BSs. Step 3 determines the set of BSs that cannot be operational in the shared band with the given transmit power range. Step 4 determines the sets of  BSs eligible to be operated in the shared band. In Step 6, for a given transmit power, we first define the following two sets, namely, $\mathbf{U}_{BS}$ and $\mathbf{U}_{sector}^{(k)}$ providing the sets of BSs and sectors to be activated. In Step 7, a score is computed for each combination of the  BSs and sectors in the $\mathbf{U}_{BS}$ and $\mathbf{U}_{sector}^{(k)}$ sets, respectively, sub-array, and beam codebook. This score provides a ratio of the overall band utilization to the (anticipated) interference (computed using the interference tool), and accordingly, a high score implies a high priority of activating a particular combination of BS, sector, sub-array, and beam codebook. In Steps 9 and 10, the combination of BS, sector, sub-array, and beam codebook with the highest priority score is selected, and the new value of interference is anticipated, respectively. If the anticipated interference exceeds the tolerable threshold, it implies that no more BS/sector can be activated with the current power, and hence,  the inner loop is terminated (Steps 11-12). Otherwise, the best combination of BS, sector, sub-array,  beam codebook, and transmit power are sequentially activated (Steps 14-29). At the end of each activation, both $\mathbf{U}_{BS}$ and $\mathbf{U}_{sector}^{(k)}$ sets are updated (Steps 15 and  24). The inner loop is continued until the $\mathbf{U}_{BS}$ set becomes empty. Meanwhile, the outer loop keeps track of the number of BSs whose transmit powers are changed in sequential outer loop iterations (Steps 17 and  26). The outer loop is terminated when all the possible transmit power values are visited or the active BSs' transmit powers can no longer be updated in consecutive iterations (Steps 31-34).

\begin{proposition}
    The worst-case time complexity of Algorithm \ref{Algorithm1} is $\mathcal{O}\left(\Delta MNK^2\right)$ where $\Delta$ is the total number of discrete transmit power levels in the $[P_{min}, P_{max}]$ range.
\end{proposition}

 \proof The proof is omitted due to space limitations.

\vspace{-0.2cm}
\section{Performance Evaluation}
\setlength{\textfloatsep}{0pt}
     \begin{figure}
     \vspace{-0.2cm}
    	\begin{center}
    	\includegraphics[width=9cm]{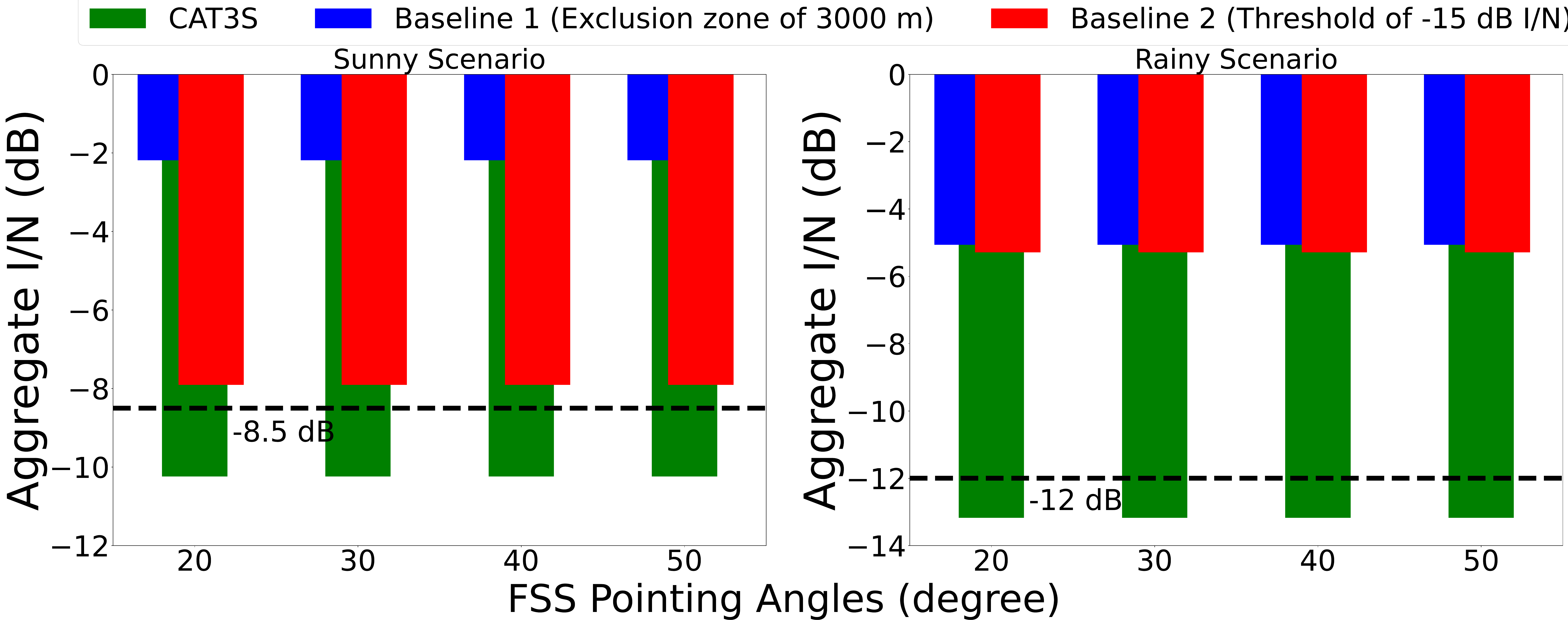}
    		\caption{Aggregate I/N ratio vs FSS's pointing angles.}
    		\label{fig_3}
    	\end{center}
     \vspace{-0.2cm}
    \end{figure}
    
\setlength{\textfloatsep}{0pt}
 \begin{figure}
    	\begin{center}
     \vspace{-0.2cm}
    	\includegraphics[width=9cm]{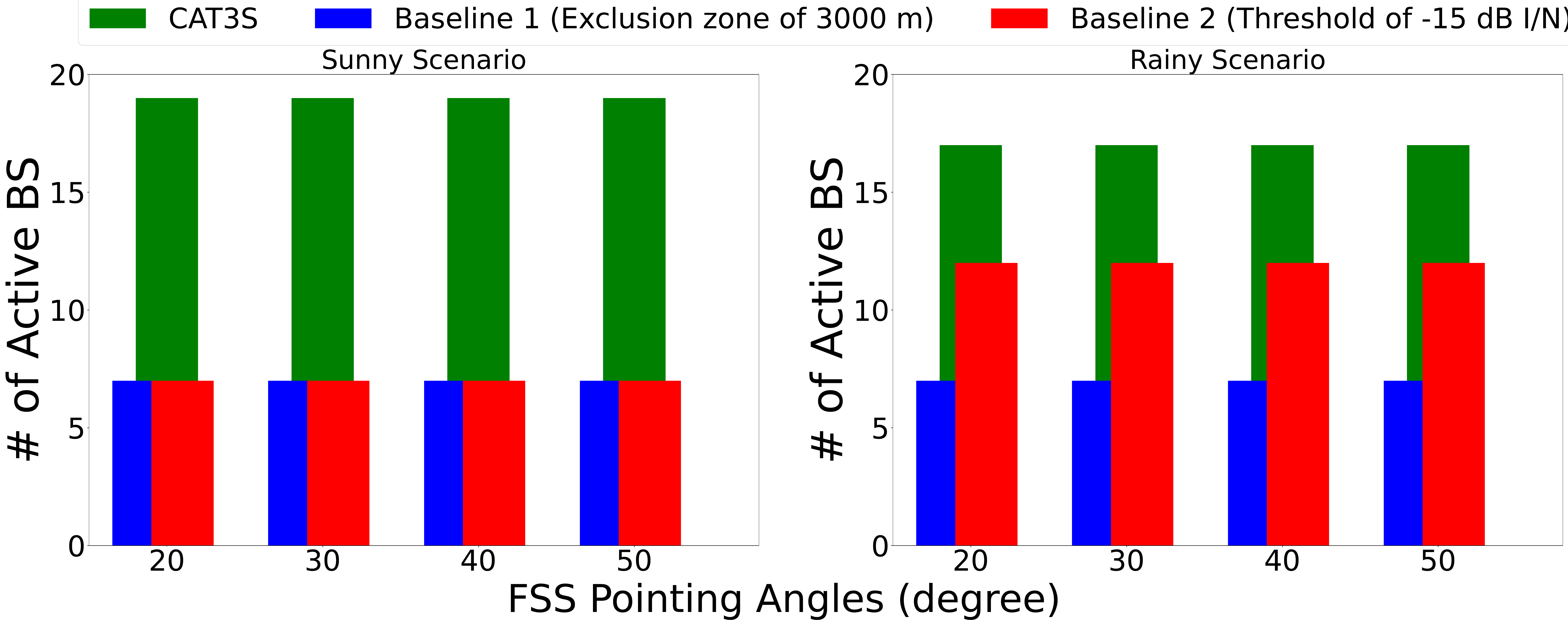}
    		\caption{ Number of active BSs vs FSS's elevation angles.}
    		\label{fig_4}
    	\end{center}
     \vspace{-0.2cm}
    \end{figure}

\subsection{Descriptions of Experimental Setup and Baseline Schemes}
%\textcolor{blue}{We have to provide the information of the Blacksburg environment, similar to WCL paper. Additional information, such as code-book-based beamforming, the FSS receiver's pointing angle should be included.}
For performance evaluation of the \proposed framework, we consider spectrum coexistence between 5G terrestrial and NGSO FSS networks over the 12 GHz band. Specifically, we exploit a real-world sub-urban deployment scenario at Blacksburg City, Virginia, from our prior study \cite{12GHz_WCL}.  In the considered deployment scenario, an FSS receiver is located at 1770 Forecast Drive in Blacksburg (37° 12' 9" North latitude and 80° 26' 4" West longitude). In addition, we incorporate actual geolocations of 33 macro BS (MBSs) from the OpenCellID database \cite{OpenCellId} within the $5000$m radius of the FSS. The heights of MBS are set to $25$ meters for the sub-urban scenario according to 3GPP specifications. Each MBS operates on a $1000$m cellular coverage area divided into three sectors ($120^\circ$ each), accommodating $10$ randomly located UEs and a total of $12$ subarrays. For an accurate and site-specific path loss analysis as discussed in Section IV, information such as heights ($10$m to $40$m), sizes, polygons, and locations of 8664 buildings is incorporated from the OpenStreetMap via overpass-turbo \cite{OpenStreetMap}. Furthermore, for the beamforming capabilities in Section IV, we consider $4\times4$ directional planner antenna array with $0.5\lambda$ element spacing and $64$ predefined beams in each subarray.  As per 3GPP, the specified nominal transmit power for an MBS is set at $-38$ dBm, and for controlling the MBS's transmit power using Algorithm \ref{Algorithm1}, we explore a power range from $-2$ dBm to $+2$ dBm relative to this nominal value. Additionally, we consider various FSS pointing angles ($20^\circ$, $30^\circ$, $40^\circ$, and $50^\circ$) and a down-tilt angle of $10^\circ$ for the MBSs, as suggested by \cite{RKFreport}. We consider weather as the specific context since the 12 GHz band's propagation characteristics vary in different weather conditions. We consider sunny and rainy weather, and these scenarios are determined based on data (e.g., the daily rain rate of a particular city) obtained from OpenWeatherMap \cite{OpenWeathertMap}. Following \cite{RKFreport}, the tolerable I/N threshold at the FSS receiver during the sunny weather is set as $-8.5$ dB. Meanwhile, due to the extensive rain fade, the received signal strength at the FSS receiver is reduced during rainy weather, and consequently, the FSS receiver requires more interference protection in rainy weather. Considering such a fact, the tolerable interference threshold at the FSS receiver during rainy weather is set as $-12$ dB.  

For performance comparison, we consider the following two baseline approaches.

\begin{itemize}
    \item \textbf{Baseline 1:} A deterministic exclusion zone of 3000m around the FSS receiver is considered. All the MBSs that are inside the exclusion zone and that are outside the exclusion zone but generate a higher I/N ratio than the tolerable threshold (i.e., $-8.5$ dB and $-12$ dB for the sunny and rainy weather conditions, respectively) are turned off. The selected BSs transmit using the nominal transmit power, and the active beams of these MBSs are selected using the conventional codebook-based beamforming algorithm.
    
    \item \textbf{Baseline 2}: In this scheme, the I/N ratio from each MBS is evaluated using our realistic interference evaluation framework, and only the MBSs generating I/N ratio smaller than a threshold remain active. Inspired by our prior study \cite{12GHz_WCL}, we consider -15 dBm as the I/N threshold for activating MBSs in the Baseline 2 scheme. The selected BSs transmit using their nominal transmit power, and the active beams of these BSs are selected using the conventional codebook-based beamforming algorithm.

\end{itemize}
%The performance between the \proposed and baseline schemes is compared in terms of the received aggregate I/N ratio at the FSS receiver, achievable downlink sum capacity, and the number of operational MBSs for different FSS receiver's pointing angles and different antenna configurations of the MBSs.

\setlength{\textfloatsep}{0pt}
           \begin{figure}
           \vspace{-0.2cm}
    	\begin{center}
    	\includegraphics[width=9cm]{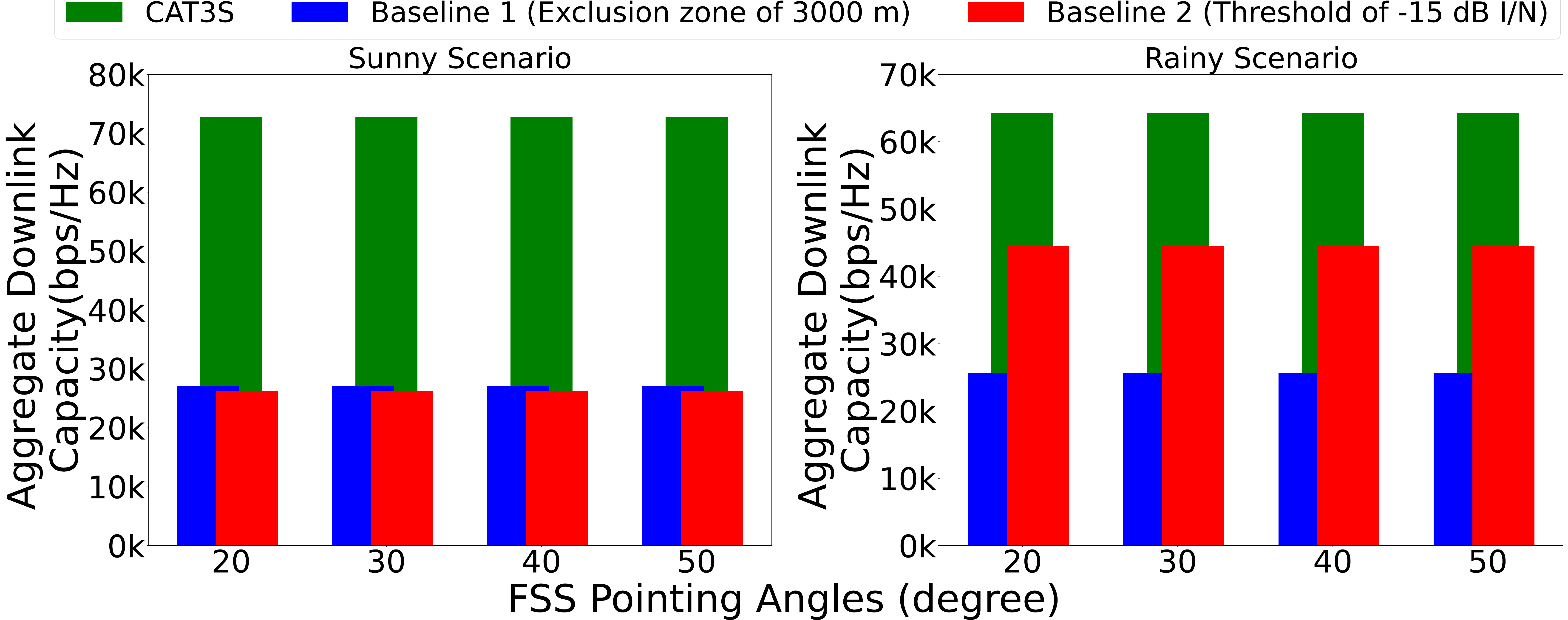}
    		\caption{Achievable downlink capacity vs. FSS pointing angles.}
    		\label{fig_5}
    	\end{center}
     \vspace{-0.2cm}
    \end{figure} 

\setlength{\textfloatsep}{0pt}
    \begin{figure}
    \vspace{-0.2cm}
    	\begin{center}
    	\includegraphics[width=9cm]{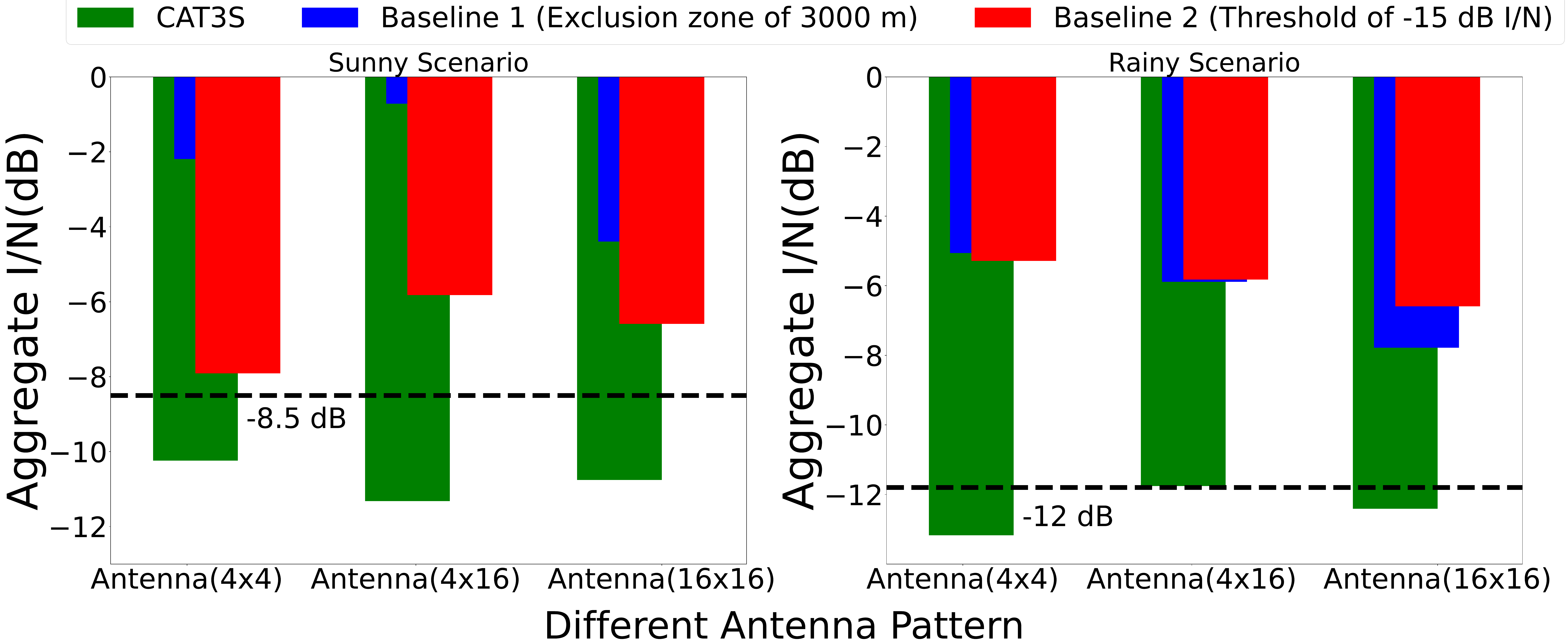}
    		\caption{ Aggregate I/N ratio vs. antenna element sizes.}
    		\label{fig_6}
    	\end{center}
     \vspace{-0.2cm}
    \end{figure}

\vspace{-0.2cm}
\subsection{Experiment Results}

\subsubsection{Performance for FSS Receiver's Various Pointing Angles}
    %variations for different pointing angles' of FSS receivers under different weather conditions. for proposed and exclusion zone-based BS on/off scheme at different 
 Figs. \ref{fig_3}, \ref{fig_4}, and \ref{fig_5} illustrate the aggregate I/N ratio received at the FSS receiver, the count of operational MBSs, and the aggregate downlink throughput for all the operational MBSs, respectively. All these performance metrics are evaluated for different FSS pointing angles in sunny (left) and rainy (right) atmospheric conditions for the \proposed and baseline schemes. For sunny and rainy weather conditions, \proposed achieves a notably smaller aggregate I/N ratio compared to both baseline schemes. For example, for $30^\circ$ pointing angle and sunny weather conditions, the \proposed, Baseline 1, and Baseline 2 schemes achieve $-10$ dBm,  $-2$ dBm, and $-8$ dBm, aggregate I/N ratios, respectively. Similarly,   for $30^\circ$ pointing angle and rainy weather conditions, the \proposed, Baseline 1 and Baseline 2 schemes achieve $-12$ dBm, $-4$ dBm, and $-4$ dBm aggregate I/N ratios, respectively. At the same time, Figs.  \ref{fig_4} depict that the \proposed keeps more MBSs operational in the shared band compared to both baseline schemes. For example, the \proposed activates $19$ and $17$ MBSs in sunny and clear weather conditions, respectively. In contrast, both baseline schemes activate $7$ MBSs in sunny weather, and for rainy weather, Baseline 1 and 2 schemes activate  $7$ and $12$ MBSs, respectively. Intuitively, the increased number of operational MBSs significantly enhances the overall achievable downlink throughput of the \proposed framework. Fig. \ref{fig_5} depicts that the overall achievable downlink throughput of \proposed reaches $75k$ bps/Hz and $65k$ bps/Hz during sunny and rainy weather, respectively. In contrast, both baseline schemes achieve $\sim 25k$ bps/Hz downlink sum capacity during the sunny weather, and during the rainy weather,  Baseline 1 and Baseline 2 schemes achieve $\sim 25k$ and $\sim 45k$   bps/Hz downlink sum capacity, respectively. The aforementioned results depict the efficacy of our \proposed framework in maximizing shared band utilization for cellular networks without harming incumbent operations in different weather conditions. We also emphasize that considered performance metrics remain the same for a wide range of FSS receiver's pointing angles. Thus,   \proposed's performance is not significantly impacted by the pointing angle selected by the FSS receiver, allowing the FSS receiver to choose the best pointing angles for its use cases. 
 
 %It is observed that for a given FSS pointing angle, the aggregated I/N values, number of operational MBSs, and the throughput of the operational MBSs derived by \proposed outperform the baseline values. Hence \proposed optimizes the parameters in such a way that the aggregate I/N always remains less than the threshold and allows the activation of a maximum number of MBSs ($19$ MBSs in sunny weather and $17$ MBSs in rainy weather). This number exceeds the count of MBSs for the two $baseline$ approaches. The active MBSs for $Baseline 1$ and $Baseline 2$ are $7$ and $7$ for sunny and $7$ and $12$ for rainy weather, respectively [Fig.\ref{fig_4}]. Meanwhile, increasing the number of operational MBSs through the \proposed algorithm significantly enhances the overall achievable downlink throughput reaching almost $75000$ bps/Hz for sunny weather and $65000$ bps/Hz while it's a rainy day [Fig. \ref{fig_5}]. This emphasizes the efficacy of our heuristic algorithm in maximizing the network's capacity by leveraging a larger number of MBSs.These values remain consistent across all the FSS elevation angles. The potential reason behind that in the simulation environment, \hl{the elevation angles used by the FSS receiver and the elevation angles employed in the optimization algorithm are identical}. 

        \setlength{\textfloatsep}{0pt}
\begin{figure}
\vspace{-0.2cm}
    	\begin{center}
    	\includegraphics[width=9cm]{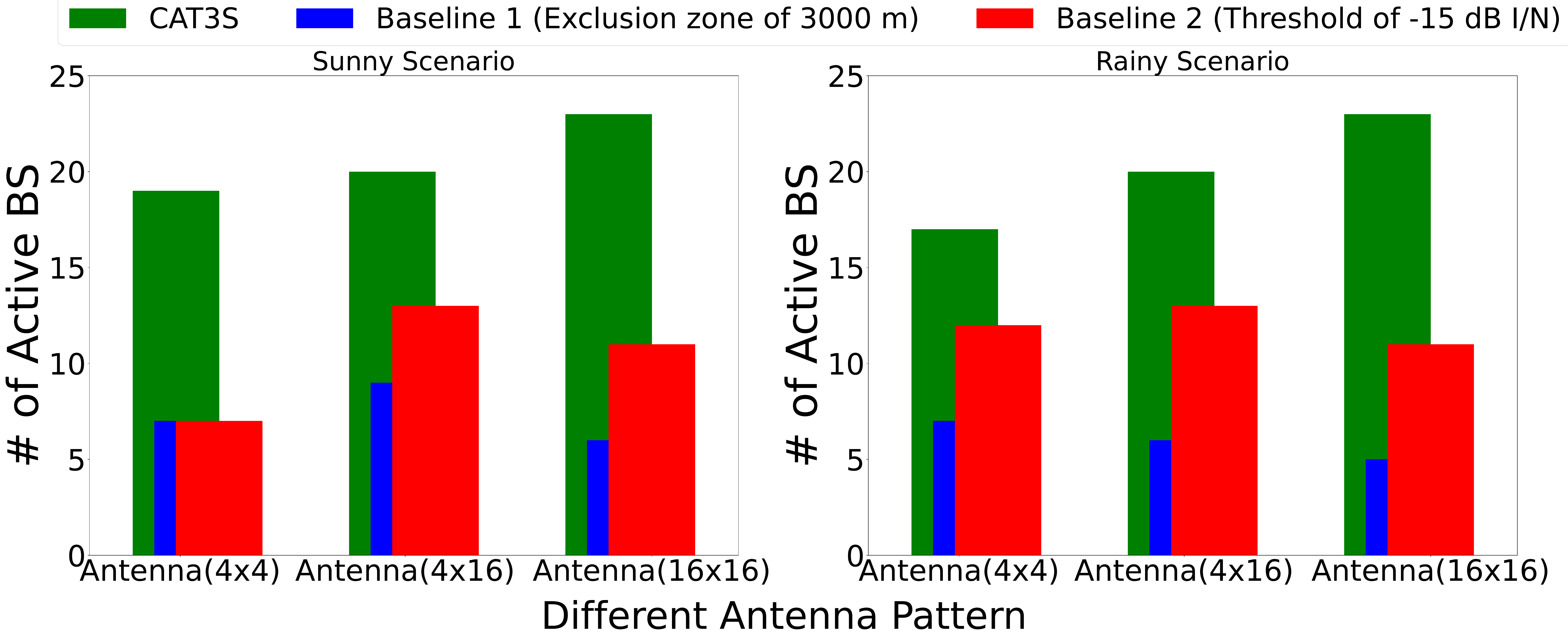}
    		\caption{ Number of active MBSs vs antenna element sizes.}
    		\label{fig_7}
    	\end{center}
     \vspace{-0.2cm}
    \end{figure}

\setlength{\textfloatsep}{0pt}
 \begin{figure}
 \vspace{-0.2cm}
    	\begin{center}
    	\includegraphics[width=9cm]{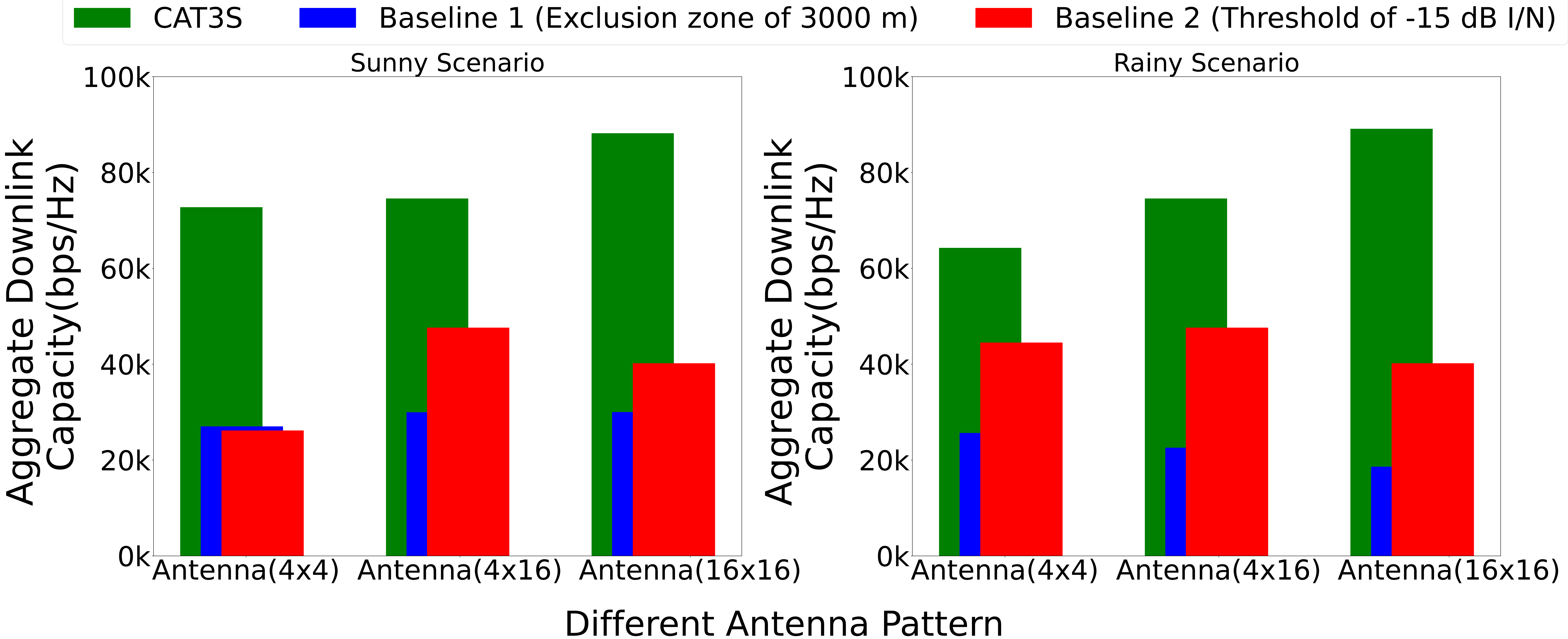}
    		\caption{Achievable downlink capacity vs. antenna element sizes.}
    		\label{fig_8}
    	\end{center}
     \vspace{-0.2cm}
    \end{figure}

% \subsubsection{Number of active MBSs for varying FSS receiver's pointing angles}
%Number of active MBSs for different pointing angles' of FSS receivers for proposed and exclusion zone-based BS on/off scheme at different weather conditions

% \subsubsection{Achievable downlink capacity for varying FSS's elevation angles}

%We present the bar graph as a comparison of our algorithm's performances with the benchmark schemes. 
% \hl{Both the benchmark schemes and proposed framework exhibit a trade-off between the number of active MBSs and achievable downlink throughput. Increasing the number of operational MBS through the BS control algorithm significantly enhances the overall achievable downlink throughput reaching almost $75000$ bps/Hz for sunny weather and $65000$ bps/Hz while it's a rainy day. However, the proposed algorithm allows the activation of the maximum number of MBSs, resulting in a higher aggregate capacity that surpasses the other two cases, which involve a lower number of active MBSs and a lower aggregate throughput. This emphasizes the efficacy of our heuristic algorithm in maximizing the network's capacity by leveraging a larger number of MBSs.}

% \subsubsection{Aggregate I/N ratio for various BS antenna array sizes}

\subsubsection{Performance for MBS's Different Antenna Array Configurations}
Figs. \ref{fig_6}, \ref{fig_7}, and \ref{fig_8} depict the variations in the aggregate I/N ratios at the FSS receiver, number of operational MBSs, and achievable aggregate downlink for different antenna array configurations at the MBSs ($4$x$4$, $4$x$16$, and $16$x$16$ in both the horizontal and vertical directions) under sunny (left) and rainy (right) weather conditions. These 
figures indicate that for all the antenna array configurations, our \proposed consistently surpasses the baseline methods by achieving a lower aggregate I/N ratio, activating more operational 5G MBSs, and enhancing system capacity. Moreover,  \proposed ensures that the aggregate $I/N$ maintains the threshold for both weather conditions. We emphasize that the \proposed selects the best set of MBSs with the combination of the optimal transmit power and the most suitable set of beams in all three sectors per MBS. Such a capability makes the \proposed more efficient than the baseline schemes.

We also observe from Figs. \ref{fig_6}-\ref{fig_8} that the aggregate I/N ratio of the \proposed does not significantly vary with the antenna array configurations. However, as the antenna array size increases, the number of operation MBSs and downlink sum capacity of the \proposed increases. This is because, with the increasing number of antenna sizes, relatively sharper and highly directional beams are obtained, which enhances the SNR and, eventually, the total system capacity. However, it is found that both baseline schemes achieve better performance for the $4 \times16$ antenna array configuration than the $16 \times16$ antenna array configuration. The potential reason is as follows. For both baselines, the suitable beams per sector and MBS are selected to maximize the downlink cellular user's performance without considering how much interference this beam could introduce at the FSS. Our experiments found that more MBSs with the $16 \times 16$ antenna array generate higher than threshold interference levels at the FSS receiver than the MBSs with the $4 \times 16$ antenna array configuration. As a result, to meet the interference threshold at the FSS receiver, for both baseline schemes, fewer MBS are operational with the $16 \times 16$ antenna array configuration, and such a fact also reduces the achievable downlink sum capacity of both baseline schemes.

\vspace{-0.2cm}
\section{Conclusion}
A context-aware dynamic spectrum sharing framework by exploiting the beamforming capability of multi-antenna terrestrial 5G MBSs, called \proposed, was proposed to enable spectrum coexistence between the terrestrial cellular and incumbent networks over the satellite bands. \proposed acquires and analyzes contexts using a context-acquisition unit and dynamically optimizes the multi-antenna MBSs’ parameters using a polynomial complexity MBS control algorithm. Such a capability enables the coexisting cellular network to activate more MBSs in the shared band without creating harmful interference at the incumbent receiver(s). For a proof-of-concept, we evaluated the \proposed framework's performance using a realistic terrestrial-FSS spectrum sharing simulator over the 12 GHz band for sub-urban deployment in Blacksburg, VA while considering weather as the context variable. Our experimental results show that in both sunny and rainy weather conditions \proposed (1) keeps the aggregate I/N ratio at the FSS receiver smaller than the tolerable thresholds and (2) achieves higher spectrum utilization for the terrestrial cellular networks than the state-of-the-art spectrum sharing approaches.

\vspace{-0.2cm}
\section*{Acknowledgement}
This work was supported by NSF grants CNS-2128540 and CNS-2128584 and by the Commonwealth Cyber Initiative (CCI), an investment by the Commonwealth of Virginia in the advancement of cyber R\&D, innovation, and workforce development.

\end{document}